\documentstyle[12pt]{article}
\begin{document}
\noindent
{\large \bf On the Source and Location of Temporal Variability in Fireballs}

\medskip
\noindent
{\bf Nir J. Shaviv }

\medskip
\noindent
{Department of Physics, Israel Institute of Technology, Haifa 32000,
Israel.}

\medskip
{\bf
\noindent
Most cosmological models for $\gamma$-ray bursts (GRBs) invoke the
production of a ``fireball'' $^{1-3}$ in a compact region, as indicated by
the short time variability of the observed GRBs. The high density of
$e^+e^-$ pairs in such fireballs inevitably makes them opaque to
$\gamma$-rays and requires the $\gamma$-ray emission to take place only
after the fireball has expanded and became optically thin to $\gamma$'s.
Many of the ``traditional'' scenarios explain the temporal variability of
GRBs as being formed by growing instabilities in the expanding
fireball$^{4}$. Here we explore this possiblity and examine its validity.}

Piran et al.$^{5}$ have shown that generally, after a short rearrangement
phase in relativistic fireballs, most of the matter and energy are
concentrated in a narrow shell. Because of the relativistic beaming, only
a small angular part of the shell with an angular size of $\theta\approx
1/\gamma$ is seen by a distant observer, where $\gamma =
1/\sqrt{1-v^2/c^2}$ is the Lorentz factor of the shell.

One can divide the possible sources of variability into three categories.
The source may be either intrinsic, in the shell, or external. In the
first case, the observed flux variability in GRBs is formed by the
fluctuations of the fireball progenitor itself. This means that the
fluctuations should have typical time scales of milliseconds, and total
durations extending upto hundreds of seconds. In the second class, the
variability is formed from instabilities in the expanding shell itself,
forming an inhomogeneous configuration that emitts $\gamma$-rays unevenly.
In the last category, the variability is formed by the interaction of the
fireball with an inhomogeneous surrounding.

Let us assume first that instabilities in the shell are the source of the
highly variable time profiles observed in GRBs. Due to the their shapes,
having long correlation lengths (and not being for example ``white
noise''$^{6}$), it is clear that the observed details originate from a
causually connected region. The typical temperature at which the bulk of
the $\gamma$-rays is emitted is less than 50 keV, at which the opacity is
reduced enough for the optical depth to decrease below unity. At these
temperatures, the largest distance a perturbation can grow relative to a
fixed point in the moving shell is: \begin{equation}
	\Delta R_{radial} = \pm {u_{thermal}\over c}{R\over \gamma^2} \approx
	0.5 {R \over \gamma^2}
	\label{}
\end{equation}
in the radial direction. Here  $R$ denotes the distance traversed by the
shell, while $u_{thermal} \approx 0.5 c$ is the typical thermal velocity
of the electrons. Similarly, in the perpendicular direction:
\begin{equation}
	\Delta R_{angular}  \approx 0.5 {R \over \gamma}.
	\label{} \end{equation}
Both these lengths correspond to durations of $ 0.5 R/c\gamma^2$ for a
distant observer.  Note that even if the electrons had highly relativistic
thermal velocities, the size would only double.  In most scenarios, the
$\gamma$-rays will be emitted from a radial region of size $\sim R$,
giving a temporal smearing over a duration of approximately $R/c\gamma^2$
in the observer frame.  This duration is larger than the possible longest
details from instabilities, making them unable to render features in the
GRBs.  Instabilities could still explain the flux variabiliy if the
emission is limited to radial scales much less then $R$ (e.g., as can
result by an improbable {\em standing} shock wave).  However, the angular
size $\theta \approx 1/\gamma$ seen by the observer must contain several
``causually disconnected regions'', since we have seen that causality
limits the size of the connected regions to an angular size smaller than
the size seen by an outside observer.  This necessarily requiers any
angular variability to come from the entire angular region.  In such a
case, the early parts of a GRB come from the central region while the
latter parts from an angle $\theta \propto \sqrt{t}$.  This would
essentially mean that typical time scales of variability increases with
time (since we ``sample'' similar features at larger time intervals - $dt
\propto \theta d\theta$).  Such a general behavior where the width of the
peaks in the temporal profile are proportional to the square root of the
time, is not seen.

It is thus a general conclusion that ``fireball'' scenarios cannot give
rise to variable GRB profiles by just utilizing internal instabilities
formed in the fireball.  Such variabilities can still arise if either the
source itself is variable on the relevant time scales, or if the mechanism
forming the $\gamma$-rays is external and nonuniform. As for the first
possibility, it is hard to constuct a model that releases cataclysmic
amounts of energy on very short times scales over durations of up to
hundreds of seconds. Therefore the later option seems the more natural way
for the temporal variability to emerge from relatevistic fireballs. This
can occur for example, when a fireball interacts with an irregular$^{7}$
interstellar medium$^{8}$. It should be stressed however that
instabilities in the fireball can manifest themselves in the GRB
light-curves, if and only if they are coupled to external inhomogeneities.

\medskip
{\small

{\bf References}

\begin{enumerate}
\item Paczy\'nski, B. {\it Astrophys. J.} {\bf 308}, L43-L46 (1986).
\item Goodman, J. {\it Astrophys. J.} {\bf 308}, L47-L50 (1986).
\item Goodman, J., Dar, A. \& Nussinov, S. {\it Astrophys. J.} {\bf 314},
L7-L10 (1987).
\item Waxmann, E. \& Piran, T., {\it Astrophys. J.} {\bf 433},
L85-L88 (1994).
\item Piran, T., Shemi A. \& Narayan R. {\it Mon. Not. R. astr. Soc.} {\bf
263}, 861-867 (1993).
\item Link, B., Epstein, R.I., \& Priedhorsky W.C. {\it Astrophys. J.}
{\bf 408}, L81-84 (1993).
\item Katz, J.I.  {\it Astrophys. J.} {\bf 422},
248-259 (1994).
\item Rees, M.I. \& M\'esz\'aros P. {\it Mon. Not. R. astr. Soc.} {\bf
258}, 41p-43p (1992).
\end{enumerate}

\noindent
 ACKNOWLEDGMENTS.
The Author wishes to thank Arnon Dar, Shlomi Pistiner, and Giora Shaviv
for fruitful discussions.}

\end{document}